\documentclass[11pt,twoside
]{article}

\usepackage{baaa2009}
\usepackage{graphicx}
\usepackage{subfigure}
\usepackage{psfrag}
\usepackage{amssymb}
\usepackage[spanish,activeacute,english]{babel}
\usepackage[latin1]{inputenc}
\usepackage[T1]{fontenc} 
\usepackage{ae,aecompl} 
\usepackage{latexsym}
\usepackage{verbatim}
\usepackage{amsmath}
\usepackage{amsfonts}
\usepackage{amssymb}
\usepackage{wasysym}
\usepackage[colorlinks=true,dvips]{hyperref}

\begin{document}
\myselectenglish
\vskip 1.0cm
\markboth{ Gustavo E. Romero }%
{The anisotropy of time}

\pagestyle{myheadings}
\vspace*{0.5cm}
\noindent TRABAJO INVITADO 
\vskip 0.3cm
\title{The anisotropy of time and the dynamics of the Universe}


\author{Gustavo E. Romero$^{1,2}$}

\affil{  
  (1) Instituto Argentino de Radioastronomía (IAR), CCT La Plata, CONICET\\
  (2) Facultad de Ciencias Astronómicas y Geofísicas - UNLP\\
}

\begin{abstract} It is a remarkable fact that all processes occurring in the observable Universe are irreversible, whereas the equations in which the laws of physics are formulated are invariant under time reversal. The emergency of irreversibility from the fundamental laws has been a topic of consideration by physicists, astronomers and philosophers since Boltzmann's formulation of the ``H'' theorem. In this paper I shall discuss several aspects of this problem and its connection with the dynamics of space-time, within the framework of modern cosmology. I shall comment, briefly, on the nature of physical time and its relation with the so-called ``consciousness''.
\end{abstract}

\begin{resumen}
  
  Es un hecho notorio que los procesos que ocurren en el universo observable son irreversibles,
mientras que las ecuaciones que expresan las leyes fundamentales de la física son invariantes bajo
inversión temporal. La emergencia de la irreversibilidad a partir de la física fundamental ha sido
un tema que ha preocupado a físicos, astrónomos y filósofos desde que Boltzmann formulara su
famoso teorema ``H''. En este trabajo discutiré varios aspectos relacionados con este problema y la
conexión del mismo con la dinámica de espacio-tiempo, en el marco de la cosmología contemporánea. 
Asimismo, comentaré brevemente sobre la naturaleza del tiempo físico y su relación con lo que se suele llamar
``conciencia''.
  
\end{resumen}

\vskip 20pt

\begin{flushright}
We shall not cease from exploration\\
And the end of all our exploring\\
Will be to arrive where we started\\
And know the place for the first time\\
\vspace{0.3cm}
T.S. Eliot\\
\vspace{0.7cm}
El rio del tiempo no desemboca en el mar de Manrique: desemboca en el efímero presente, en el aquí y ahora de este línea que está corriendo, que usted está leyendo, y que tras sus ojos se está yendo conmigo hacia la nada.\\
\vspace{0.3cm}
Fernando Vallejo

\end{flushright}

\section{Introduction}

There is something notorious about the world around us. It {\em changes}. Moreover, on the long run, the sense of such changes seems not to be very promising. As Pascal wrote: 
\begin{quotation}
The last act is tragic, however happy all the rest of the play is; at the last a little earth is thrown upon our head, and that is the end for ever\footnote{{\em Pensées}, Random House, New York, 1941, p.45}.
\end{quotation}
Less tragically, the past seems to be quite different from the future. We can remember the former and, sometimes, predict the latter. We grow older, not younger. The Universe was hotter in the past, and very likely it will become colder in the future. The disorder around us seems to increase. All these facts and many others of the kind are expressed in terms of the so-called Second Law of Thermodynamics: {\em The entropy of a closed system never decreases}. If entropy is denoted by $S$, this law reads:
\begin{equation}
	\frac{dS}{dt}\geq 0.
\end{equation}

In the 1870s, Ludwig Boltzmann (1844-1906) argued that the effect of randomly moving gas molecules was to ensure that the entropy of the gas would increase, until it reached its maximum possible value. This is his famous {\em H-theorem}. Boltzmann was able to show that macroscopic distributions of great inhomogeneity (i.e. of high order or low entropy) are formed from relatively few microstate arrangements of molecules, and were, consequently, relatively improbable. Since physical systems do not tend to go into states that are less probable than the states they are in, it follows that any system would evolve toward the macrostate that is consistent with the larger number of microstates. The number of microstates and the entropy of the system are related by the fundamental formula:
\begin{equation}
	S= k \ln W,
\end{equation}
where $k=10^{23}$ JK$^{-1}$ is Boltzmann's constant and $W$ is the volume of the phase-space that corresponds to the macrostate of entropy $S$. 

More than twenty years after the publication of Boltzmann's fundamental papers on kinetic theory (Boltzmann 1872, 1877), it was pointed out by Burbury (1894, 1895) that the source of asymmetry in the H-theorem is the assumption that the motions of the gas molecules are independent before they collide and not afterward, if entropy is going to increase. This essentially means that the entropy increase is a consequence of the {\em initial conditions} imposed upon the state of the system. Boltzmann's response was (Boltzmann 1895): 

\begin{quotation}
There must then be in the universe, 
which is in thermal equilibrium as a 
whole and therefore dead, here and 
there, relatively small regions of the 
size of our  world, which during the 
relatively short time of eons deviate 
significantly from thermal equilibrium.  
Among these worlds the state probability 
increases as often as it decreases.        
\end{quotation} 

As noted by Price (2004): ``The low-entropy condition of our region seems to be associated entirely with a low-energy condition in our past.''

The probability of the large fluctuations required for the formation of the Universe we see, by other hand, seems to be computable in zero, as noted long ago by Eddington (1931): ``A universe containing mathematical physicists 
 at any assigned date will be in the state of 
maximum disorganization which is not inconsistent 
with the existence of such creatures.'' Large fluctuations are rare ($P\sim \exp{-\Delta S}$); {\em extremely} large fluctuation, basically impossible . For the whole universe, $\Delta S\sim 10^{104}$ in units of $k=1$ (Egan \& Lineweaver 2009). This yields $P=0$. However, we are here, living because we are, momentously,  far from thermal equilibrium.    

\section{Formulation of the problem}

In 1876, a former teacher of Boltzmann and later colleague at the University of Vienna, J. Loschmidt, noted:

\begin{quotation}
Obviously, in every arbitrary system the course of events must become retrograde when the velocities of all its elements are reversed. (Loschmidt 1876).
\end{quotation}

Putting the point in modern terminology, the laws of (Hamiltonian) mechanics are such that for every solution one can construct another solution by reversing all velocities and replacing $t$ by $-t$. Since the Boltzmann's function $H[f]$ is invariant under the velocity reversal, it follows that if $H[f]$ decreases for the first solution, it will increase for the second. Accordingly, the reversibility objection is that the H-theorem cannot be a general theorem for all mechanical evolutions of the gas. More generally, the problem goes far beyond classical mechanics and encompasses our whole representation of the physical world. This is because {\em all formal representations of all fundamental laws of physics are invariant under the operation of time reversal}. Nonetheless, the evolution of all physical processes in the Universe is irreversible. 

If we accept, as mentioned in the introduction, that the origin of the irreversibility is not in the laws but in the initial conditions of the laws, two additonal problems emerge: 1) what were exactly these initial conditions?, and 2) how the initial conditions, of global nature, can enforce, at any time and any place, the observed local irreversibility?. 

The first problem is, in turn, related to the following one, once the cosmological setting is taken into account: in the past, the Universe was hotter and at some point matter and radiation were in thermal equilibrium; how is this compatible with the fact that entropy has ever been increasing according to the so-called Past Hypothesis, i.e. entropy was at a minimum at some past time and has been increasing ever since?.  

The standard answer to this question invokes the expansion of the Universe: as the Universe expanded, the maximum possible entropy increased with the size of the Universe, but the actual entropy was left well behind the permitted maximum. The Second Law of Thermodynamics and the source of irreversibility is the trend of the entropy to reach the permitted maximum. According to this view, the Universe actually began in a state of maximum entropy, but due to the expansion, it was still possible for the entropy to continue growing.        

The main problem with this line of thought is that is not true that the Universe was in a state of maximum disorder at some early time. In fact, although locally matter and radiation might have been in thermal equilibrium, this situation occurred in a regime were the global effects of gravity cannot be ignored (Penrose 1979). Since gravity is an attractive force, and the Universe was extremely smooth (i.e structureless) in early times, as indicated, for instance, by the measurements of the cosmic microwave background radiation, the gravitational field should have been quite far from equilibrium, with very low global entropy (Penrose 1979). It seems, then, that the early Universe was {\em globally} out of the equilibrium, being the total entropy dominated by the entropy of the gravitational field. If we want to understand the implications of this, first we need to briefly discuss the concepts of space and time, and their relation to gravitation.     


\section{What is space-time?}
 
In the Summer of 1908, Hermann Minkowski (1864-1909) was at the peak of his creativity. During September of that year, he presented his view of the electrodynamics of moving bodies at the annual meeting of the Society of German Scientists and Physicians, held in Cologne. The title of his talk was ``Space and Time'', and his opening words have became justly famous (Minkowski 1908):
\begin{quotation}
The views of space and time which I wish to lay before you have sprung from the soil of experimental physics, and therein lies their strength. They are radical. Henceforth space by itself, and time by itself, are doomed to fade away into mere shadows, and only a kind of union of the two will preserve an independent reality.
\end{quotation}   

The concept of space-time introduced by Minkowski would play a key role in the formulation of a new view of gravitation, when it was applied by Albert Einstein (1879-1955) in the framework of his search for a General Theory of Relativity. What is this space-time of Minkowski and Einstein?. \\

{\em Space-time is the ontological sum of all events of all things.}\\

A thing is an individual endowed with physical properties. An event is a change in the properties of a thing. An ontological sum is an aggregation of things or physical properties, i.e. a physical entity or an emergent property. An ontological sum should not be confused with a set, which is a mathematical construct and has only mathematical (i.e. fictional) properties (see Pérez \& Romero 2009 for more details). 

Everything that has happened, everything that happens, everything that will happen, is just an element, a ``point'', of space-time. Space-time is not a thing, it is just the relational property of all things. 

As it happens with every physical property, we can represent space-time with some mathematical structure, in order to describe it. We shall adopt the following mathematical structure for space-time:\\

{\sl Space-time can be represented by a differentiable, 4-dimensional, real manifold.}\\ 

If we want to calculate distances between two events, we need more structure on our manifold: we need a geometric structure. We can introduce such an structure with a metric tensor $g_{\mu\nu}$ that tell us how to calculate distances. The distance $ds$ between two arbitrarily close events is: 
\begin{equation}
ds^{2}=g_{\mu\nu} dx^{\mu} dx^{\nu}.
\end{equation}

\begin{figure}[!ht]
  \centering
  \includegraphics[width=.65\textwidth]{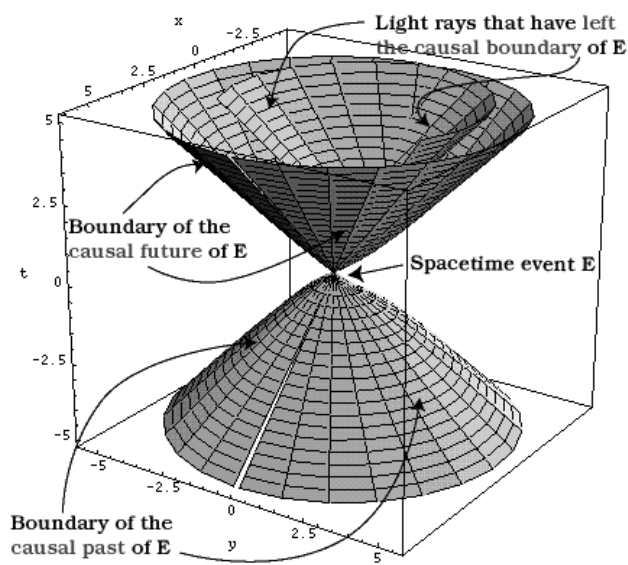}
 \caption{{\it Light cone in Minkowskian (flat) space-time}} 
  \label{fig:Romero1}
\end{figure}

Once we have introduced a metric tensor we can separate space-time at each point in three regions according to $ds^{2}<0$ (space-like region), $ds^{2}=0$ (light-like or null region), and $ds^{2}>0$ (time-like region). Particles that go through the origin can only reach time-like regions. The null surface $ds^{2}=0$ can be inhabited only by particles moving at the speed of light, like photons. Points in the space-like region cannot been reached by material objects from the origin of the {\sl light cone} that can be formed at any space-time point. Notice that the introduction of the metric allows to define the future and the past {of a given event}. Let us consider the unitary vector $T^{\nu}=(1,\;0,\;0,\;0)$, then a vector $x^{\mu}$ points to the future if $\eta_{\mu\nu}   x^{\mu} T^{\nu}>0$. In similar way, the vector points toward the past if $\eta_{\mu\nu}   x^{\mu} T^{\nu}<0$. A light cone is shown in Figure \ref{fig:Romero1}.

\section{The present}

Events in space-time do not flow. They simply {\em are}. What is called the transient ``now'' or ``present'' is not itself an event in space-time. Events are ordered by the relations ``earlier than'' or ``later than'', but no event is singled out as ``present'', except by convention. As noted by Adolf Gümbaum (1963), what is ``now'' about a given event is that it is affecting some conscious being who is aware of the event. The ``present'' is not an intrinsic property of a given event, much less a changing thing in the world, but a relation among some number of events and a self-conscious individual.  More specifically, we can define: \\

{\em Present: class of all events that are physically related to given brain event.}\\

The present, consequently, like smell, sweetness, and other secondary qualities, is introduced by the interaction of sentient individuals with their environment. Time does not ``flow'' in any physical sense. What changes is the state of our consciousness of the events in our surroundings. In words of Hermann Weyl (1885-1955):

\begin{quotation}
	The objective world simply is, it does not happen. Only the gaze of my consciousness, crawling upward along the life line of my body, does a section of this world come to life as a fleeting image in space which continuously  changes in time (Weyl 1949). 
\end{quotation}

The ``becoming'' is not a property of the events of space-time, but a property of our consciousness of such events. We call ``becoming'' to the series of states of consciousness associated with a series of physical events. Time does not go by. We do.   

\section{Curvature and time-orientability}

A general manifold is not necessary flat. The metric tensor can be a function of the coordinates of space-time points. If such a general manifold $(M,\;g_{\mu\nu})$ has at any point a tangent space with a Lorentzian metric (trace $-2$), we say that the represented space-time is {\em pseudo-Rimaniann}. In a general space-time the curvature is determined through the Riemann tensor:
\begin{equation}
	R^{\sigma}_{\mu\nu\lambda}=\partial_{\nu}\Gamma^{\sigma}_{\mu\lambda}-\partial_{\lambda}\Gamma^{\sigma}_{\mu\nu}
	+\Gamma^{\sigma}_{\alpha\nu}\Gamma^{\alpha}_{\mu\lambda}-\Gamma^{\sigma}_{\alpha\lambda}\Gamma^{\alpha}_{\mu\nu},\label{Riemann}
\end{equation} 
where $\Gamma^{\lambda}_{\mu\lambda}$ is the affine connection of space-time, related to the metric by:
\begin{equation}
\Gamma^{\lambda}_{\mu\nu}=\frac{1}{2}g^{\lambda\alpha}(\partial_{\mu}g_{\nu\alpha}+\partial_{\nu}g_{\mu\alpha}-\partial_{\alpha}g_{\mu\nu}).
\end{equation}

The metric represents the gravitational potential, whereas the affine connection the gravitational field. 

We say that a given space-time $(M, \; g_{\mu\nu})$ is {\em time-orientable} if we can define over $M$ a smooth non-vanishing time-like vector field. Notice that space-times with closed time-like curves or black holes are in general not time-orientable. Time, being represented by a one-dimensional continuum, cannot be anisotropic. What is anisotropic is the time-like vector field, which essentially indicates the direction of strings of events (physical processes) in the general manifold. It is just in such a case that, by a language abuse, we can talk of the ``direction'' of time, or the ``anisotropy'' of time. The expression ``the arrow of time'', coined by Eddington, is essentially incorrect, since time is {\em not} a vector\footnote{Some authors confuse the anisotropy of the physical processes in space-time with an intrinsic ``time-asymmetry". Time itself is not time-asymmetric; what are asymmetric are processes in space-time. See Horwich (1987) for a discussion.}.   

\section{Electrodynamics and pseudo-Riemannian space-time}

The electromagnetic radiation can be described in the terms of the 4-potential $A^{\mu}$, which in the Lorentz gauge satisfies:
\begin{equation}
	\partial^{\nu}\partial_{\nu}A^{\mu}(\vec{r},\;t)=4\pi j^{\mu} (\vec{r},\;t),
\end{equation}
with $c=1$ and $j^{\mu}$ the 4-current. With appropriate boundary conditions is possible to write $A^{\mu}$ as a functional of the sources $j^{\mu}$. The retarded and advanced solutions are:

\begin{equation}
	A^{\mu}_{\rm ret}(\vec{r},\;t)=\int
\frac{j^{\mu} \left(\vec{r},\;t-\left|\vec{r}-\vec{r'}\right|\right)}{\left|\vec{r}-\vec{r'}\right|}d^{3}\vec{r'}, \label{ret}
\end{equation}

\begin{equation}
	A^{\mu}_{\rm adv}(\vec{r},\;t)=\int
\frac{j^{\mu} \left(\vec{r},\;t+\left|\vec{r}-\vec{r'}\right|\right)}{\left|\vec{r}-\vec{r'}\right|}d^{3}\vec{r'}. \label{adv}
\end{equation}

The two functionals of $j^{\mu}(\vec{r},\;t)$ are related to one another by a time reversal transformation. The solution (\ref{ret}) is contributed by all sources in the causal past of the space-time point $(\vec{r},\;t)$ and the solution (\ref{adv}) by all the sources in the causal future of that point. The linear combinations of these solutions are also solutions, since the equations are linear and the Principle of Superposition holds. It is usual to consider only the retarded potential as physical meaningful in order to estimate the electromagnetic field at $(\vec{r},\;t)$: $F^{\mu\nu}_{\rm ret}=\partial^{\mu}A^{\nu}_{\rm ret}-\partial^{\nu}A^{\mu}_{\rm ret}$. However, there seems to be no compelling reason for such a choice. We can adopt, for instance, (in what follows we use a simplified notation and boundary conditions such as the surface contribution is zero):
\begin{equation}
	A(\vec{r},\;t)=\frac{1}{2}\int_{V} ({\rm adv\; + \; ret})\; dV.
\end{equation}

\begin{figure}[!ht]
  \centering
  \includegraphics[width=.45\textwidth]{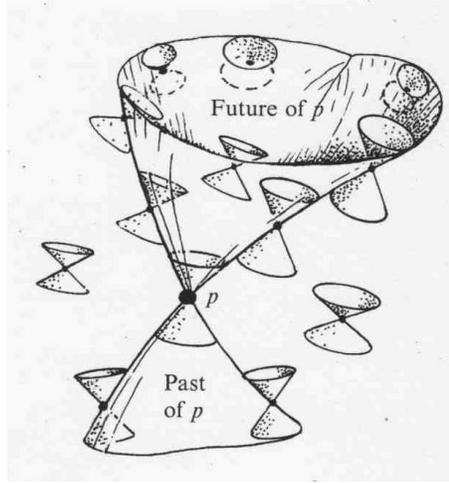}
 \caption{{\it Light cones in general pseudo-Riemannian space-time (after Penrose 2007).}} 
  \label{fig:Romero2}
\end{figure}
 
If space-time is curved ($R\neq 0$), the null cones that determine the local causal structure will not be symmetric around the point $p(\vec{r},\;t)$, see Fig. \ref{fig:Romero2}. Then, 
\begin{equation}
L=\lim_{V \to \infty} \left[\int_{V} \rm adv - \int_{V} \rm ret \right]\; dV \neq 0.
\end{equation}

If $L\neq0$ there is a preferred direction for the flow of the Poynting flux in space-time. In a black hole interior this direction is always toward the singularity. In an expanding Universe, it is in the global future direction. We see, then, that time (time-like physical processes), in a general space-time $(M, g_{ab})$, is (are) {\sl anisotropic}. There is a global to local relation given by the Poynting flux as determined by the curvature of space-time that indicates the direction in which events occur. Time-like processes, inside a black hole, have a different orientation from outside. The causal structure of the world is determined by the dynamics of space-time and the initial conditions. 

Notice that the electromagnetic flux is related to the macroscopic concept of temperature through the Stefan-Boltzmann law: $$L=A\;\sigma_{\rm SB}T^{4},$$ 
where $\sigma_{\rm SB}= 5.670 400 \times 10^{-8} {\rm J\,s}^{-1} {\rm m}^{-2} {\rm K}^{-4}$ is the 
Stefan-Boltzmann constant. From this abd the preceding considerations it follows the Second Law of Thermodynamics.

Macroscopic irreversibility and time anisotropy, then, emerge from essentially reversible laws. This will happen even in space-times with no global time orientability.

\section{The entropy of gravitation}

The entropy of gravitation plays a crucial role in the specification of the initial conditions of the Universe. If gravitation can be quantified as other fields, then we might expect to have an entropy associated with a gas of gravitons. A macroscopic state of space-time, represented by a given geometry (metric) should correspond to many gravitational microstates. In the absence of a quantum theory of gravitation, a statistical approach is inviable. Penrose (1979), then, has suggested that the Weyl curvature tensor can be used to specify the gravitational entropy. The Weyl tensor is a 4-rank tensor that contains the independent components of the Riemann tensor not captured by the Ricci tensor. It can be considered as the traceless part of the Riemann tensor (\ref{Riemann}).
In dimensions 2 and 3 the Weyl curvature tensor vanishes identically. In $n\geq 4$ dimensions, the Weyl curvature is generally nonzero. 
If the Weyl tensor vanishes, then there exists a coordinate system in which the metric tensor is proportional to a constant tensor. 

The Weyl tensor can be obtained from the full curvature tensor by subtracting out various traces. The Riemann tensor has 20 independent components, 10 of which are given by the Ricci tensor and the remaining 10 by the Weyl tensor.  

The Weyl tensor is given in components by
\begin{equation}
C_{\alpha\beta\gamma\delta}=R_{\alpha\beta\gamma\delta}+\frac{2}{n-2}(g_{\alpha[\gamma}R_{\delta]\beta}-g_{\beta[\gamma}R_{\delta]\alpha})+\frac{2}{(n-1)(n-2)}R~g_{\alpha[\gamma}g_{\delta]\beta},
\end{equation}
where $R_{\alpha\beta\gamma\delta}$ is the Riemann tensor, $R_{\alpha\beta}$ is the Ricci tensor, $R$ is the Ricci scalar and $[]$ refers to the antisymmetric part. In 4 dimensions the Weyl tensor is:
\begin{equation}
	C_{\alpha\beta\gamma\delta}=R_{\alpha\beta\gamma\delta}+\frac{1}{2}\left(g_{\alpha\delta}R_{\gamma\beta}+g_{\beta\gamma}R_{\delta\alpha}-g_{\alpha\gamma}R_{\delta\beta}-g_{\beta\delta}R_{\gamma\alpha}\right)+\frac{1}{6}\left(g_{\alpha\gamma}g_{\delta\beta}-g_{\alpha\delta}g_{\gamma\beta}\right)R.
\end{equation}

Since entropy is a scalar, in the case of a gravitational field is expected that it should be measured by some scalar constructed out of a full contraction of the Weyl tensor. The simplest candidate is: 
\begin{equation}
	C^{2}=C^{\alpha\beta\gamma\delta}C_{\alpha\beta\gamma\delta}.
\end{equation}
Other possibilities, like
\begin{equation}
	C^{2}=\frac{C^{\alpha\beta\gamma\delta}C_{\alpha\beta\gamma\delta}}{R^{\alpha\beta}R_{\alpha\beta}},
\end{equation}
or
\begin{equation}
	C^{2}=\frac{C^{\alpha\beta\gamma\delta}C_{\alpha\beta\gamma\delta}}{R^{\alpha\beta\gamma\delta}R_{\alpha\beta\gamma\delta}},
\end{equation}    
have been proposed as well. 

The initial conditions from which physical irreversibility arise then require that $C^{2}\sim 0$ in the very early Universe.  


\section{Causal explanations?}

Do the initial conditions of the Universe, namely the fact that $C^{2}\sim 0$, require a causal explanation?. What such an explanation would be?.  

The causal relation is a relation between events (ordered pairs of states), not between things. Causation is a form of event generation (Pérez \& Romero 2009). The initial conditions represent a state of a thing (the Universe in this case, the maximal thing) and hence have no causal power. The initial conditions are a ``state of affairs''. The causal power should be looked for in previous events, but if space-time itself, as an emergent property of basic things, has a quantum behavior, classical causality would not operate. Rather, the initial conditions should appear as a classical limit of the gravitational processes at quantum level. Final conditions can be causally explained because there are possible causes that precede final conditions, but initial conditions, on the contrary, cannot be causally explained because there are no times that precede them. The initial conditions of the Universe, then, should have an explanation in terms of yet unknown dynamical laws. Such laws do not need, and likely have not, a causal structure.

\section{Final remarks}

Time is an emergent property of changing things. It is represented by a one-dimensional continuum. Processes in space-time are anisotropic, although physical laws are invariant under time reversal. Time itself is not anisotropic, because it is not represented by a vector field. The dynamical state of space-time and the initial conditions determine the local direction of the physical processes through the electromagnetic Poynting flux. There is a global-to-local relation between gravitation and electrodynamics that determines the increase of entropy. The present is not an event. The present is not a feature of a world without consciousness. It is a secondary quality, like the smell of a rainy day or the sadness of a love gone. 

\begin{flushright}
\begin{verse}
	Arrojado a quietud,\\
divisaré esa playa última de tu ser\\
y te veré por vez primera, quizá,\\
como Dios ha de verte,\\
desbaratada la ficción del Tiempo\\
\vspace{0.5cm}
\hspace{4.5cm} Borges
\end{verse}
\end{flushright}


\begin{referencias}

\reference Burbury, S.H. 1894, Nature, 51, 78
\reference Burbury, S.H. 1895, Nature, 51, 320
\reference Boltzmann, L. 1872, Wiener Berichte, 66, 275
\reference Boltzmann, L. 1877, Wiener Berichte, 76, 373
\reference Boltzmann, L. 1895, Nature, 51, 413    
\reference Eddington, A.S. 1931, Nature, 127, 3203                                                                            
\reference Egan, C.A., \& Lineweaver, C.H. 2009, \apj, submitted (arXiv:0909.3983)
\reference Grümbaum, A. 1963, Philosophical Problems of Space and Time, A. Knopf, New York
\reference Horwich, P. 1987, Asymmetries in Time, The MIT Press, Cambridge 
\reference Loschmidt, J. 1876, Wiener Berichte, 73, 128 
\reference Minkowski, H. 1908, in: The Principle of Relativity, Dover, New York, 1952, p.75 
\reference Penrose, R. 1979, in: General Relativity: An Einstein Centenial, S.W. Hawking \& W. Israel (eds.), Cambridge University Press, Cambridge, p. 581
\reference Penrose, R. 2007, The Road to Reality, Vintage Books, New York 
\reference Pérez, D., \& Romero, G.E. 2009, BAAA 52, in press
\reference Price, H. 2004, in: Contemporary Debates in Philosophy of Science, C. Hitchcock (ed.), Blackwell, Singapore, p. 219  
\reference Weyl, H. 1949, Philosophy of Mathematics and Natural Science, Princeton University Press, Princeton
\end{referencias}

\end{document}